\documentclass[twocolumn,showpacs,preprintnumbers,amsmath,amssymb]{revtex4-1}
\usepackage{bbm}
\usepackage{amsfonts}

\usepackage{epsfig}
\usepackage{graphicx}
\usepackage{amssymb}   
\usepackage{dcolumn}
\usepackage{bm}
\usepackage[colorlinks,citecolor=blue, linkcolor=blue, urlcolor=blue, hyperindex]{hyperref}
\begin{document}
\vspace{2.5cm}
\title{Enhanced quantum teleportation in the background of Schwarzschild spacetime by weak measurements}
\author{Xing Xiao$^{1}$}
\author{Yao Yao$^{2}$}
\author{Yan-Ling Li$^{3}$}
\altaffiliation{liyanling0423@gmail.com}
\author{Ying-Mao Xie$^{1}$}
\altaffiliation{xieyingmao@126.com}
\affiliation{$^{1}$College of Physics and Electronic Information, Gannan Normal University, Ganzhou, Jiangxi 341000, China\\
$^{2}$Microsystems and Terahertz Research Center, China Academy of Engineering Physics, Chengdu, Sichuan 610200, China\\
$^{3}$School of Information Engineering, Jiangxi University of Science and Technology, Ganzhou, Jiangxi 341000, P. R. China}

\begin{abstract}
It is commonly believed that the fidelity of quantum teleportation in the gravitational field would be degraded due to the heat up by the Hawking radiation. In this paper, we point out that the Hawking effect could be eliminated by the combined action of pre- and post-weak measurements, and thus the teleportation fidelity is almost completely protected. It is intriguing to notice that the enhancement of fidelity could not be attributed to the improvement of entanglement, but rather to the probabilistic nature of weak measurements.
Our work extends the ability of weak measurements as a quantum technique to battle against gravitational decoherence in relativistic quantum information.
\end{abstract}
\maketitle

\section{Introduction}
\label{intro}

Relativistic quantum information is becoming a rapidly developing new field of physics in recent years \cite{mann2012}. It focuses on the key problem that how the relativistic effects impact on the entanglement and entanglement-based quantum information tasks \cite{schuller2005,bradler2009,fuentes2010,emm2010a,emm2010b,bruschi2010,yao2014,bruschi2014,ahmadi2014,dai2015,wang2016}.
A deeper comprehension of the relativistic quantum information not only plays a crucial role in understanding the non-relativistic quantum information processing \cite{peres2004}, but also sheds light on the black hole thermodynamics and information paradox of black holes \cite{hawking1974,zhang2009,zhang2013}. In addition to these theoretical interests, another exciting development in this field is that several quantum information experiments are already reaching relativistic regimes \cite{ursin2007,wilson2011,rideout2012,vallone2016}.

Quantum teleportation is one of the most interesting implications of the `weirdness' of quantum
mechanics \cite{bennett1993}. It is a fundamental ingredient of the long-distance quantum communication and quantum networks \cite{nielsen2000}. In the past few decades, outstanding progresses have been achieved in both theoretical and experimental study of quantum teleportation \cite{bouwmeester1997,furusawa1998,braunstein1998,kim2001,riebe2004,barret2004,olms2009,pfaff2014,wang2015}. Particularly, over 100-kilometer free-space channel quantum teleportation has been realized in the experiments \cite{yin2012}. The forthcoming satellite-based teleportation naturally inspires researchers to consider the relativistic effects in quantum teleportation. Alsing and Milburn demonstrated that the fidelity of the teleportation between an inertial and an {\it uniformly} accelerated observer is reduced due to the Unruh radiation \cite{alsing2003,alsing2004}. Then it was extended to the case of continuous variables and {\it nonuniformly} accelerated observer \cite{friis2013}. Additionally, the teleportations in the gravitational fields, such as in the background of Schwarzschild spacetime and in an expanding space \cite{ge2005,pan2007,feng2013}, are also examined. Almost without exception, the fidelity of teleportation in these scenarios is greatly degraded due to the Unruh-Hawking radiation. Although numerous researches have demonstrated the fidelity degradation in noninertial frames or curved spacetime, there is still rare study devoted to protecting the fidelity of teleportation against the Unruh-Hawking radiation.

In this paper, we study the quantum teleportation of Dirac fields which are observed by Alice and Bob in the background of Schwarzschild spacetime. In order to ensure the teleportation could be achieved, Bob moves with uniform acceleration and hovers near the event horizon of the Schwarzschild black hole. Different from the preceding teleportation protocols \cite{alsing2003,alsing2004,ge2005,pan2007,feng2013,land2009}, two weak measurements are involved before and after Bob suffers the Hawking radiation, which are called as pre- and post-weak measurements, respectively. These two weak measurements aim to eliminate the influence of Hawking radiation. Namely, a pre-weak measurement is intentionally performed to move Bob's state close to the `lethargic' state which is immune to the Hawking radiation. 
Therefore the collapsed state is maintained for a long time with little deterioration. Finally, the post-weak measurement is performed to remove the effects of pre-weak measurement and Hawking radiation, and thus restore the teleported state. We find that the assistance of 
weak measurements enables the fidelity to be almost completely recovered to 1, thereby nullifying the decoherence effect of Hawking radiation. We also propose two types of optimal post-weak measurements that both drastically improve the teleportation fidelity. It is intriguing to observe that the fidelity of teleportation could be improved even without increasing the entanglement. This mechanism is quite distinct from any conventional improvement of the teleportation fidelity through the increase of shared entanglement. 

The organization of this paper is as follows. In Sec. \ref{sec2}, we review the vacuum structure for Dirac fields in the Schwarzschild spacetime and introduce the concept of weak measurement. In Sec. \ref{sec3}, we then show how the fidelity of teleportation could be enhanced by
weak measurements. We propose two different types of optimal post-weak measurements which are both effective in combating Hawking effect. In Sec. \ref{sec4}, the underlying mechanism of our protocol is discussed in detail. Finally, the conclusions are summarized in Sec. \ref{sec5}.

\section{Preliminaries}
\label{sec2}
\subsection{Vacuum structure and Hawking radiation for Dirac fields in the Schwarzschild spacetime}
\label{sec2.1}
Here, we review the essential features of Dirac fields in the Schwarzschild spacetime. The metric for the Schwarzschild spacetime is given by (hereafter Plank units are used $\hbar=G=c=k=1$)
\begin{equation}
\label{eq1}
ds^{2}=-\left(1-\frac{2M}{r}\right)dt^{2}+\left(1-\frac{2M}{r}\right)^{-1}dr^{2}+r^{2}d\Omega^{2},
\end{equation}
where $d\Omega^{2}=d\theta^{2}+\sin^{2}\theta d\phi^{2}$. $M$ and $r$ are the mass and radius of the black hole. In the curved spacetime, the massless Dirac equation is described as \cite{birrell1984}
\begin{eqnarray}
\label{eq2}
&\frac{-\gamma_{0}}{\sqrt{1-\frac{2M}{r}}}\frac{\partial\Psi}{\partial
t}+\gamma_{1}\sqrt{1-\frac{2M}{r}}\left[\frac{\partial}{\partial r}+\frac{1}{r}+\frac{M}{2r(r-2M)}\right]\Psi \nonumber\\
&+\frac{\gamma_{2}}{r}\left(\frac{\partial}{\partial\theta}+\frac{1}{2}\cot\theta\right)\Psi+
\frac{\gamma_{3}}{r\sin\theta}\frac{\partial\Psi}{\partial\phi}=0,
\end{eqnarray}
where $\gamma_{i}, (i=0,1,2,3)$ are the 4 by 4 Dirac matrices.

Solving the Dirac equation (\ref{eq2}) near the event horizon $r_{h}$, we obtain the
positive frequency outgoing solutions outside (region \rm I) and inside (region \rm II) of the
event horizon, respectively.
\begin{eqnarray}
\label{eq3}
\psi_{\bf k}^{\rm I_{+}}=&&\mathcal{G}e^{-i\omega u}, (r>r_{h}), \\
\psi_{\bf k}^{\rm II_{+}}=&&\mathcal{G}e^{i\omega u},(r<r_{h}).\nonumber
\end{eqnarray}
Here $u=t-r_{*}$, $\mathcal{G}$ is a 4-component Dirac spinor and
$r_{*}=r+2M\ln\left|\frac{r-2M}{2M}\right|$ is the tortoise
coordinate. Then the Dirac field $\Psi$ can be expressed in above bases
\begin{equation}
\label{eq}
\Psi=\int d\mathbf{k}\left(a_{\mathbf{k}}^{\rm I}\psi_{\mathbf{k}}^{\rm I_{+}}+b_{\mathbf{-k}}^{\rm I\dagger}\psi_{\mathbf{k}}^{\rm I-}+a_{\mathbf{k}}^{\rm II}\psi_{\mathbf{k}}^{\rm II_{+}}+b_{\mathbf{-k}}^{\rm II\dagger}\psi_{\mathbf{k}}^{\rm II-}\right),
\end{equation}
where $a_{\mathbf{k}}^{\rm I}$, $a_{\mathbf{k}}^{\rm II}$ and $b_{\mathbf{-k}}^{\rm I\dagger}$, $b_{\mathbf{-k}}^{\rm II\dagger}$ correspond to the fermion annihilation and antifermion creation operators acting on the state in region \rm I and \rm II, respectively. 
 
In order to make the analytic continuations of Eq. (\ref{eq3}), one usually introduces the generalized light-like Kruskal coordinates
$\mathcal{U}$ and $\mathcal{V}$,
\begin{eqnarray}
\label{eq4}
&&\mathcal{U}=-\exp\left({-\frac{u}{4M}}\right), ~~~ \mathcal{V}=\exp\left({\frac{v}{4M}}\right),~\rm region ~I\\
&&\mathcal{U}=\exp\left({-\frac{u}{4M}}\right), ~~~ \mathcal{V}=-\exp\left({\frac{v}{4M}}\right), ~\rm region ~II\nonumber
\end{eqnarray}
where $u=t-r_{*}$, $v=t+r_{*}$, and $r_{*}$ is a tortoise coordinate. The regions \rm I and \rm II are denoted by
$\mathcal{U}<0$ and $\mathcal{U}>0$, respectively. Then we have the
Schwarzschild metric of Eq. (\ref{eq1}) in Kruskal coordinates,
\begin{equation}
\label{eq5}
ds^{2}=-\frac{32M^3}{r}\exp{\left(-\frac{r}{2M}\right)}d\mathcal{U}d\mathcal{V}+r^{2}d\Omega^{2}.
\end{equation}
where $r=r(\mathcal{U},\mathcal{V})$ is defined in terms of $\mathcal{U}$ and $\mathcal{V}$ by the implicit equation $\mathcal{U}\mathcal{V}=-\frac{r-2M}{2M}\exp(\frac{r}{2M})$.

According to the suggestion by Damour and Ruffini \cite{damoar1976}, it
is straightforward to find a complete basis for the positive energy
Kruskal modes,
\begin{eqnarray}
\label{eq6}
\mathcal{F}_{\mathbf{k}}^{\rm I_{+}}=&&e^{2M\pi\omega}\psi_{\mathbf{k}}^{\rm I_{+}}+e^{-2M\pi\omega}\psi_{-\mathbf{k}}^{\rm II_{-}}, \\
\mathcal{F}_{\mathbf{k}}^{\rm II_{+}}=&&e^{2M\pi\omega}\psi_{-\mathbf{k}}^{\rm I_{-}}+e^{-2M\pi\omega}\psi_{\mathbf{k}}^{\rm II_{+}}.\nonumber
\end{eqnarray}
In this new basis, we can expand the Dirac field $\Psi$ in the Kruskal spacetime
\begin{eqnarray}
\Psi=\int d\mathbf{k}&&\big[2\cosh(4M\pi\omega)\big]^{-1/2}\bigg(c_{\mathbf{k}}^{\rm I}\mathcal{F}_{\mathbf{k}}^{\rm I_{+}}+d_{\mathbf{-k}}^{\rm I\dagger}\mathcal{F}_{\mathbf{k}}^{\rm I-} \nonumber\\
&&+c_{\mathbf{k}}^{\rm II}\mathcal{F}_{\mathbf{k}}^{\rm II_{+}}+d_{\mathbf{-k}}^{\rm II\dagger}\mathcal{F}_{\mathbf{k}}^{\rm II-}\bigg),
\end{eqnarray}
where $c_{\mathbf{k}}^{\rm I}$ and $d_{\mathbf{-k}}^{\rm I\dagger}$ are the  annihilation and creation operators acting on the Kruskal vacuum.

Then we can quantize the Dirac fields in the Schwarzschild and Kruskal
modes in terms of Eqs. (\ref{eq3}) and (\ref{eq6}), respectively.
According to the Bogoliubov transformations \cite{barnett1997} between
creation and annihilation operators of the Kruskal and Schwarzschild
coordinates, we have the relevance of the vacua between the two
kinds of spacetimes. After properly normalizing the state vector, the Kruskal vacuum and the only excited states are expressed  by \cite{wang2010},
\begin{eqnarray}
\label{eq9}
|0_{\mathbf{k}}\rangle^{K}=&&\zeta|0_{\mathbf{k}}\rangle^{\rm I}|0_{-\mathbf{k}}\rangle^{\rm II}+\eta|1_{\mathbf{k}}\rangle^{\rm I}|1_{-\mathbf{k}}\rangle^{\rm II},\\
\label{eq10} 
|1_{\mathbf{k}}\rangle^{K}=&&|1_{\mathbf{k}}\rangle^{\rm I}|0_{-\mathbf{k}}\rangle^{\rm II},
\end{eqnarray}
where $\zeta=\left(e^{-\omega/T}+1\right)^{-1/2}$,
$\eta=\left(e^{\omega/T}+1\right)^{-1/2}$. $T=1/8\pi M$ is the
Hawking temperature and $|n_{-\bf k}\rangle_{\rm II}$, $|n_{\bf k}\rangle_{\rm I}$
are the orthogonal bases for the inside and outside regions of the
event horizon, respectively.

\subsection{Pre- and post-weak measurements}
\label{sec2.2}
The weak measurement used in our scheme is positive operator valued measure, which is in contrast to the standard von Neumann projective measurement that projects the initial state to one of the eigenstates of the observable. Weak measurements only barely disturb the system by state collapse, thereby retaining the measured state reversible \cite{korotkov2006,katz2008,kim2012,xiao2016}. The elements of our pre-weak measurement could be formally written as \cite{xiao2017}
\begin{eqnarray}
\label{eq11}
m_{0}&=&\sqrt{\overline{p}}|0\rangle\langle0|+|1\rangle\langle1|,\\
\label{eq12}
m_{1}&=&\sqrt{p}|0\rangle\langle0|,
\end{eqnarray}
with $\overline{p}=1-p$ and $m_{0}^{\dagger}m_{0}+m_{1}^{\dagger}m_{1}=I$. The parameter $p$ $(0\leqslant p\leqslant1)$ is usually known as the strength of weak measurement. Since $m_{1}$ is identical to the projective measurement and results in an irrevocable collapse, we discard the outcome of the measurement $m_{1}$. The successful action of measurement operator $m_{0}$ is a weak measurement, which maps the state of a qubit to an unnormalized state
\begin{eqnarray}
\label{eq13}
\rho\xrightarrow{\Lambda_{m_{0}}(\rho)}\left[\begin{array}{cc}\overline{p}\rho_{00} & \sqrt{\overline{p}}\rho_{01} \\\sqrt{\overline{p}}\rho_{10} & \rho_{11}\end{array}\right],
\end{eqnarray}
where $\rho_{ij}$ $(i,j=0,1)$ are the elements of the initial state in the computational basis. 

\begin{figure}
 \includegraphics[width=0.5\textwidth]{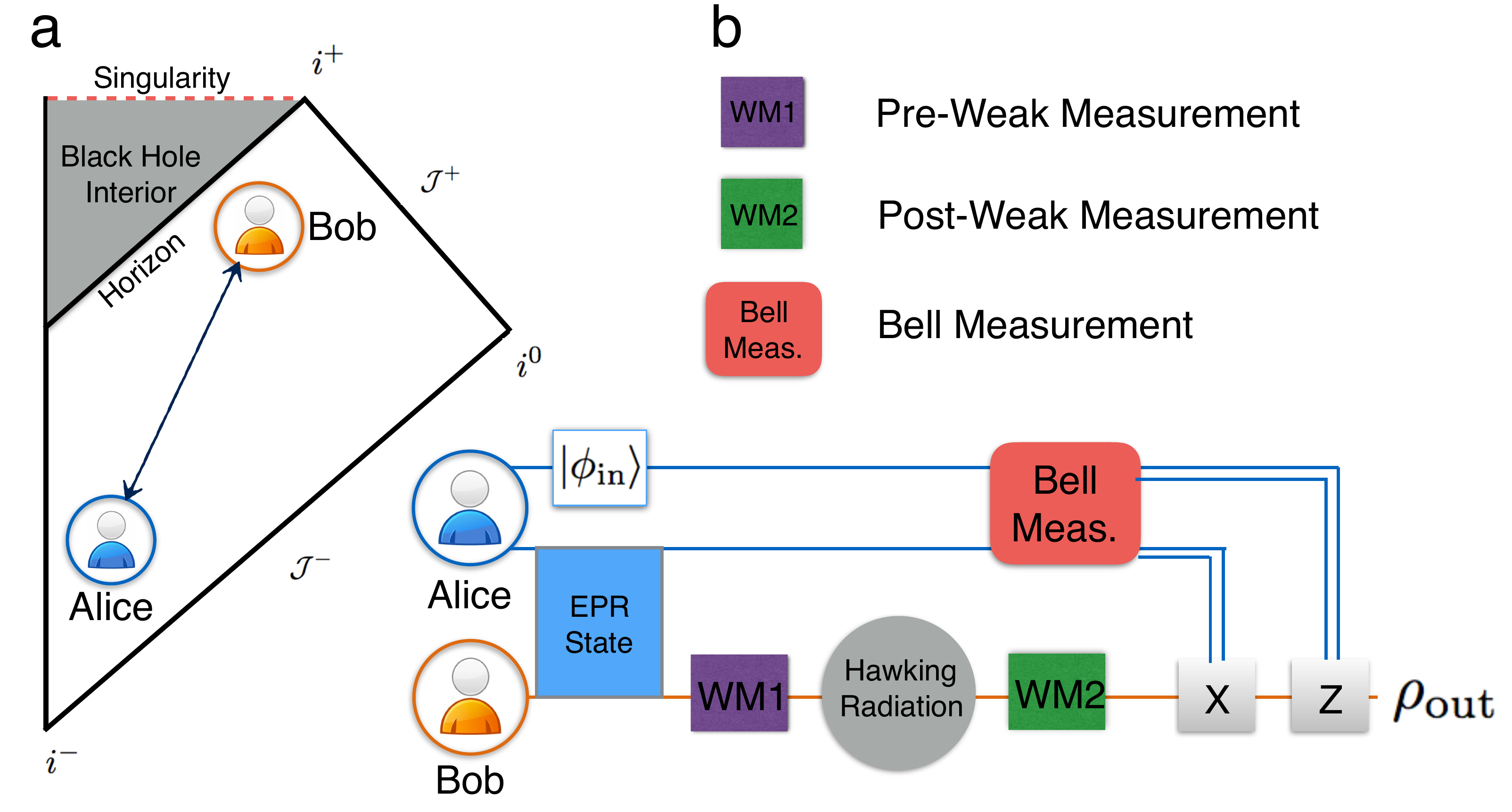}
\caption{(color online) Schematic illustrations of weak measurements enhanced quantum teleportation in the background of Schwarzschild spacetime. (a) Penrose diagram of Schwarzschild spacetime. (b) The circuit of quantum teleportation with the assistance of weak measurements.}
\label{Fig1}       
\end{figure}

Note that the discarded outcome of $m_{1}$ implies that the outcome of $m_{0}$ keeps the state $|0\rangle$ unchanged with a probability $1-p$.
Hence the initial state can be recovered with a nonzero success probability by reversing operation (i.e., a post-weak measurement) \cite{cheong2012}.
The post-weak measurement consists of a set of positive operators, such that
\begin{eqnarray}
\label{eq14}
M_{0}&=&|0\rangle\langle0|+\sqrt{\overline{q}}|1\rangle\langle1|,\\
\label{eq15}
M_{1}&=&\sqrt{q}|1\rangle\langle1|.
\end{eqnarray}
Similarly, we still give up the outcome of $M_{1}$ while keep the result of $M_{0}$. The post-weak measurement $M_{0}$ could exactly undo the pre-weak measurement $m_{0}$ in a probabilistic way by properly choosing the parameter $q$. For the ideal case without decoherence, the sequential actions of $m_{0}$ and $M_{0}$ on the qubit can be expressed as
 \begin{eqnarray}
\label{eq16}
\rho\xrightarrow{\Lambda_{M_{0}}\Lambda_{m_{0}}(\rho)}\left[\begin{array}{cc}\overline{p}\rho_{00} & \sqrt{\overline{pq}}\rho_{01} \\\sqrt{\overline{pq}}\rho_{10} & \overline{q}\rho_{11}\end{array}\right].
\end{eqnarray}
If we choose $q=p$, it is evident that the final state yields to the initial state with a constant factor $1/(1-p)$ which is related to the non-unitary nature of the weak measurement, i.e., probabilistic. On the other hand, it is interesting to note that $M_{0}$ could be re-written as $M_{0}=X\widetilde{m}_{0}X$ with $X=|0\rangle\langle1|+|1\rangle\langle0|$ the bit-flip operation (i.e., Pauli X operation). This indicates that the post-weak measurement could be constructed by three steps: a bit-flip operation, a second pre-weak measurement $\widetilde{m}_{0}$ with measurement strength $q$ and a second bit-flip operation.

\section{Enhancing the fidelity of teleporttion by weak measurements}
\label{sec3}
We are now in a position to demonstrate the basic idea of our protocol. As shown in Fig.~\ref{Fig1}a, we consider the setting that consists of two observers (Alice and Bob). They share an Einstein-Podolsky-Rosen (EPR) state at the same point in the flat region. After their coincidence, Bob freely falls into the Schwarzschild black hole and finally hovers near the event horizon, while Alice remains at the flat region. In order to enhance the fidelity of teleportation, the pre- and post-weak measurements are performed before and after the Hawking radiation, as illustrated in the Fig.~\ref{Fig1}b.

We assume the shared EPR state between Alice and Bob is $|\Psi^{+}\rangle=(|00\rangle+|11\rangle)/\sqrt{2}$ and the teleported state hold by Alice is an arbitrary superposition state
\begin{equation}
\label{eq17}
|\phi_{\rm in}\rangle=\alpha|0_{\rm in}\rangle+\beta|1_{\rm in}\rangle,
\end{equation}
where $\alpha=\cos\frac{\theta}{2}$ and $\beta=\sin\frac{\theta}{2}e^{i\delta}$ with $\theta\in[0,\pi]$ and $\delta\in[0,2\pi]$. Thus, the initial state of the whole system is
\begin{equation}
\label{eq18}
|\psi_{1}\rangle=\frac{1}{\sqrt{2}}|\phi_{\rm in}\rangle\left(|0_{\rm A}0_{\rm B}\rangle+|1_{\rm A}1_{\rm B}\rangle\right).
\end{equation}

As we demonstrated in Fig.~\ref{Fig1}b, a pre-weak measurement $m_{0}$ is carried out before Bob suffering the Hawking radiation, then the state $|\psi_{1}\rangle$ will reduce to
\begin{equation}
\label{eq19}
|\psi_{2}\rangle=\frac{1}{\sqrt{2-p}}|\phi_{\rm in}\rangle\left(\sqrt{\overline{p}}|0_{\rm A}0_{\rm B}\rangle+|1_{\rm A}1_{\rm B}\rangle\right).
\end{equation}
Note that the amplitude of the state $|1_{\rm A}1_{\rm B}\rangle$ is magnified due to the shrink of the normalization factor. As $p$ approaches unity, the EPR state is almost completely projected into $|1_{\rm A}1_{\rm B}\rangle$. This intentional projection enables Bob to be immune from the following Hawking radiation, as we will discuss below.

When Bob hovers near the event horizon of Schwarzschild black hole, he must uniformly accelerate to stay outside the horizon and suffers the Hawking radiation. Thus, the state of Bob should be specified in Kruskal coordinates in order to describe what Bob sees. Substituting the Eqs. (\ref{eq9}) and (\ref{eq10}) into $|\psi_{2}\rangle$, we arrive the state 
\begin{eqnarray}
\label{eq20}
|\psi_{3}\rangle=&&\frac{1}{\sqrt{2-p}}\bigg(\alpha\sqrt{\overline{p}}\zeta|0000\rangle+\alpha\sqrt{\overline{p}}\eta|0011\rangle+\alpha|0110\rangle \nonumber\\
&&+\beta\sqrt{\overline{p}}\zeta|1000\rangle+\beta\sqrt{\overline{p}}\eta|1011\rangle+\beta|1110\rangle\bigg).
\end{eqnarray}
For convenience, we have introduced the following notation $|ijkl\rangle=|i_{\rm in}\rangle|j_{\rm A}\rangle|k_{\rm I}\rangle|l_{\rm II}\rangle$. In the next step, a post-weak measurement is performed by Bob. Since Bob is causally disconnected from region II, the post-weak measurement only acts on the region I. Then the state changes to 
\begin{eqnarray}
\label{eq21}
|\psi_{4}\rangle=&&\frac{1}{\sqrt{N}}\bigg(\alpha\sqrt{\overline{p}}\zeta|0000\rangle+\alpha\sqrt{\overline{pq}}\eta|0011\rangle+\alpha\sqrt{\overline{q}}|0110\rangle \nonumber\\
&&+\beta\sqrt{\overline{p}}\zeta|1000\rangle+\beta\sqrt{\overline{pq}}\eta|1011\rangle+\beta\sqrt{\overline{q}}|1110\rangle\bigg).
\end{eqnarray}
where $N=\overline{p}\zeta^2+\overline{q}+\overline{pq}\eta^2$ is the normalization factor. 

\begin{figure*}[t]
 \includegraphics[width=0.9\textwidth]{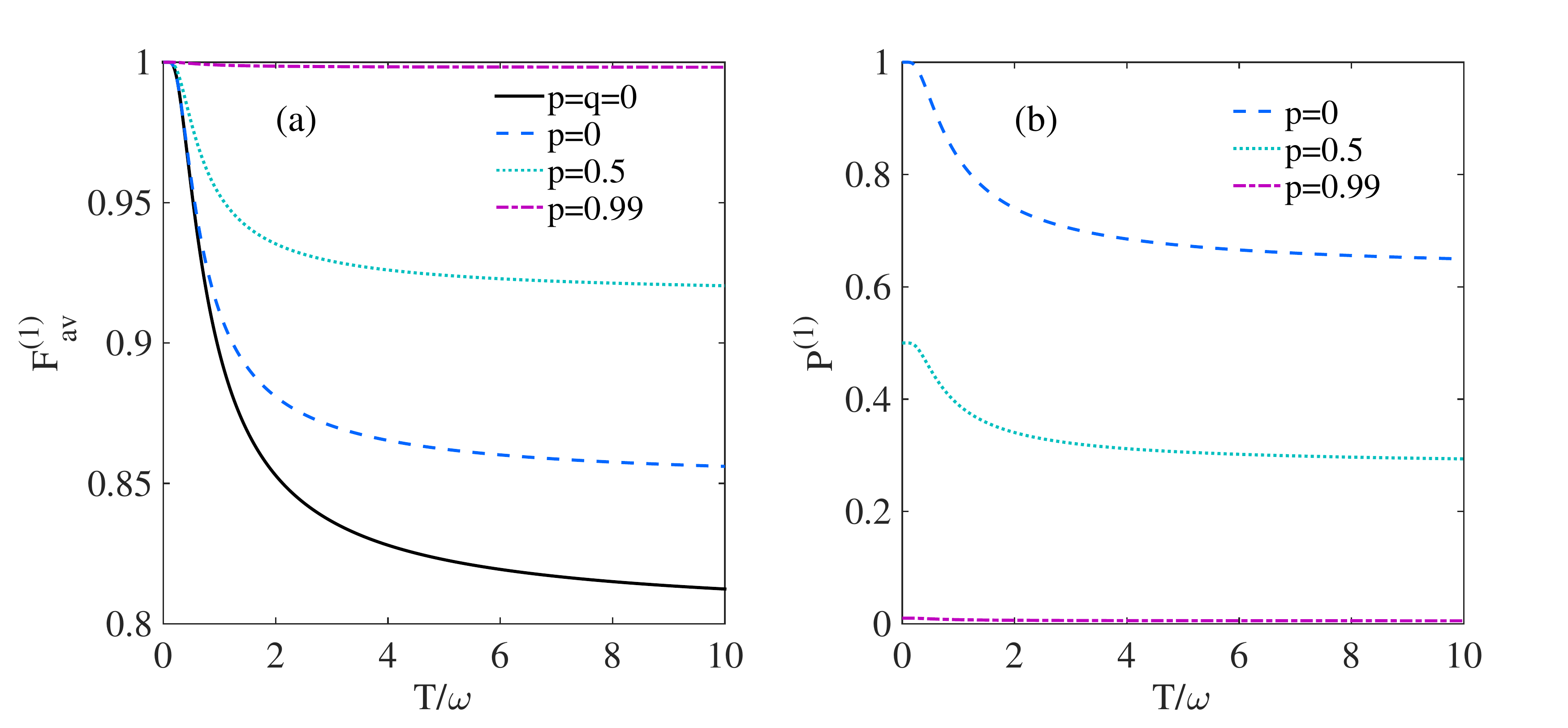}
\caption{(color online)  (a) The average fidelity $F_{\rm av}^{(1)}$ and (b) the success probability $P^{(1)}$ as a function of scaled Hawking temperature for various values of $p$ under the optimal strength of the post-weak measurement given by Eq. (\ref{eq27}).}
\label{Fig2}       
\end{figure*}

Then Alice performs a joint measurement on her two qubits and sends the results to Bob. In this process, one should note that Alice will cross Bob's future event horizon in a finite time $t_{c}$, after which the light signals cannot reach Bob. Therefore, to ensure the classical communication, Alice must send her measurement results to Bob before this critical time. 
For Bob, the observer locates near the event horizon of a black hole, he needs to trace over the modes
in the interior region since he is causally disconnected from this region. According to Alice's results, the corresponding local unitary operations are carried out in region I. A routine computation gives rise to the following result.
\begin{eqnarray}
\label{eq22}
|\psi_{5}\rangle=&&\frac{1}{\sqrt{N}}\bigg[\alpha\sqrt{\overline{p}}\zeta(|0000\rangle+|1000\rangle)+\alpha\sqrt{\overline{q}}(|0100\rangle+|1100\rangle)\nonumber\\
&&+\alpha\sqrt{\overline{pq}}\eta(|0011\rangle-|1011\rangle)+\beta\sqrt{\overline{p}}\zeta(|0110\rangle+|1110\rangle)\nonumber\\
&&+\beta\sqrt{\overline{pq}}\eta(|0101\rangle-|1101\rangle)+\beta\sqrt{\overline{q}}(|0010\rangle+|1010\rangle)\bigg].\nonumber\\
\end{eqnarray}
Consequently, the received state by Bob turns out to be
\begin{eqnarray}
\label{eq23}
\rho_{\rm out}=&&\frac{1}{N}\bigg[(\alpha^2\overline{p}\zeta^2+\alpha^2\overline{q}+|\beta|^2\eta^2\overline{pq})|0\rangle\langle0|\\
&&+2\alpha\beta^{*}\zeta\sqrt{\overline{pq}}|0\rangle\langle1|+2\alpha\beta\zeta\sqrt{\overline{pq}}|1\rangle\langle0|\nonumber\\
&&+(|\beta|^2\overline{p}\zeta^2+|\beta|^2\overline{q}+\alpha^2\eta^2\overline{pq})|1\rangle\langle1|\bigg].\nonumber
\end{eqnarray}
The fidelity of this teleportation is now given by
\begin{eqnarray}
\label{eq24}
F&&=\langle\phi_{\rm in}|\rho_{\rm out}|\phi_{\rm in}\rangle\\
&&=\frac{(\alpha^4+|\beta|^4)(\overline{p}\zeta^2+\overline{q})+2\alpha^2|\beta|^2\overline{pq}\eta^2+4\alpha^2|\beta|^2\sqrt{\overline{pq}}\zeta}{N}.\nonumber
\end{eqnarray}
Considering the teleported state $|\phi_{\rm in}\rangle$ is unknown, one should calculate the average fidelity. As the Eq. (\ref{eq23}) only involves the term $|\beta|^2$ which is independent of the phase parameter $\delta$, hence the average fidelity could be written as 
\begin{eqnarray}
\label{eq25}
F_{\rm av}&&=\frac{1}{\pi}\int_{0}^{\pi}F(\theta)d\theta\\
&&=\frac{3(\overline{p}\zeta^2+\overline{q})+\overline{pq}\eta^2+2\sqrt{\overline{pq}}\zeta}{4N}.\nonumber
\end{eqnarray}

To remove the effect of Hawking radiation and achieve the maximal average fidelity, how to choose the optimal strength of the post-weak measurement is crucial. In the following two subsection, we propose two types of post-weak measurement. 
Both of them effectively enhance the average fidelity of teleportation.

\subsection{type 1}
An intuitive method for improving the average fidelity is choosing a post-weak measurement with proper strength that ensures the final state as close as possible to the initial state. The thermal excitation induced by Hawking radiation could be regarded as a quantum transition, which results in the Kruskal vacuum state $|0_{\mathbf{k}}\rangle^{K}$ jumping to the only excited state $|1_{\mathbf{k}}\rangle^{\rm I}$ with the probability $\eta^2$. Therefore, one can technically use the mathematical trick of `unraveling' the thermal excitation into `jump' and `no jump' scenarios and work with pure states \cite{scully1997}. For this purpose, we re-express the Eq. (\ref{eq22}) as following
\begin{widetext}
\begin{eqnarray}
\label{eq26}
|\psi_{5}\rangle=&&\frac{1}{\sqrt{N}}\bigg[|00\rangle\big(\alpha\sqrt{\overline{p}}\zeta|0\rangle+\beta\sqrt{\overline{q}}|1\rangle\big)|0\rangle+|10\rangle\big(\alpha\sqrt{\overline{p}}\zeta|0\rangle+\beta\sqrt{\overline{q}}|1\rangle\big)|0\rangle+|01\rangle\big(\alpha\sqrt{\overline{q}}|0\rangle+\beta\sqrt{\overline{p}}\zeta|1\rangle\big)|0\rangle\\
&&+|11\rangle\big(\alpha\sqrt{\overline{q}}|0\rangle+\beta\sqrt{\overline{p}}\zeta|1\rangle\big)|0\rangle+\alpha\sqrt{\overline{pq}}\eta(|0011\rangle-|1011\rangle)+\beta\sqrt{\overline{pq}}\eta(|0101\rangle-|1101\rangle)\bigg].\nonumber
\end{eqnarray}
\end{widetext}
Clearly, the optimal strength of the post-weak measurement yields to
\begin{equation}
\label{eq27}
\overline{q}^{(1)}=\overline{p}\zeta^2.
\end{equation}
Under such a condition, the states possessed by Bob in the first four terms of Eq. (\ref{eq26}) both turn out to be $\overline{p}\zeta(\alpha|0_{\rm I}\rangle+\beta|1_{\rm I}\rangle)$. Then the degradation of fidelity only stems from the last two terms which consist the parameter $\eta$. The average fidelity of Eq. (\ref{eq25}) changes into
\begin{equation}
\label{eq28}
F_{\rm av}^{(1)}=\frac{8+\overline{p}\eta^2}{8+4\overline{p}\eta^2}.
\end{equation}
Intriguingly, we find this unfavorable impact of $\eta$ could be greatly reduced by increasing the measurement strength of pre-weak measurement $p$, which could be verified by setting $\overline{p}\rightarrow0$ in above equation. In order to get a distinct impression of the power of weak measurements we plot the average fidelities of teleportation as a function of Hawking temperature (scaled by $\omega$) for different strengths of pre-weak measurement in Fig.~\ref{Fig2}a.

Three remarks are in order: Firstly, in the absence of weak measurements, i.e., $p=q=0$, the average fidelity decreases as the Hawking temperature increases, which directly indicates the degradation of entanglement initially shared between Alice and Bob since the state fidelity in conventional teleportation protocol is related to the entanglement. This result is consistent with  that obtained in Refs. \cite{ge2005,pan2007,feng2013}. Secondly, the fidelity is also partially improved by the post-weak measurement even without the pre-weak measurement ($p=0$). It could be considered as a simple means of error correction based on the prior knowledge of Hawking temperature. Thirdly, the combination of pre- and post weak measurements can be indeed employed for enhancing the fidelity of teleportation in the background of gravitational field. Particularly, in the limit $p\rightarrow1$, it is clear that the effect of Hawking radiation can be circumvented by exploiting pre- and post-weak measurements simultaneously,
the fidelity hence is almost completely protected.

Since the weak measurements are not unitary, the enhancement of fidelity is not deterministic but probabilistic. Under the condition of Eq. (\ref{eq27}), the success probability is
\begin{equation}
\label{eq29}
P^{(1)}=\frac{\overline{p}\zeta^2}{2}(2+\overline{p}\eta^2).
\end{equation}
Obviously, the success probability decreases with increasing strength of pre-weak measurement, which means that the high fidelity teleportation is achieved at the expense of low success probability, as shown in Fig.~\ref{Fig2}b.

\subsection{type 2}
\begin{figure*}
 \includegraphics[width=0.9\textwidth]{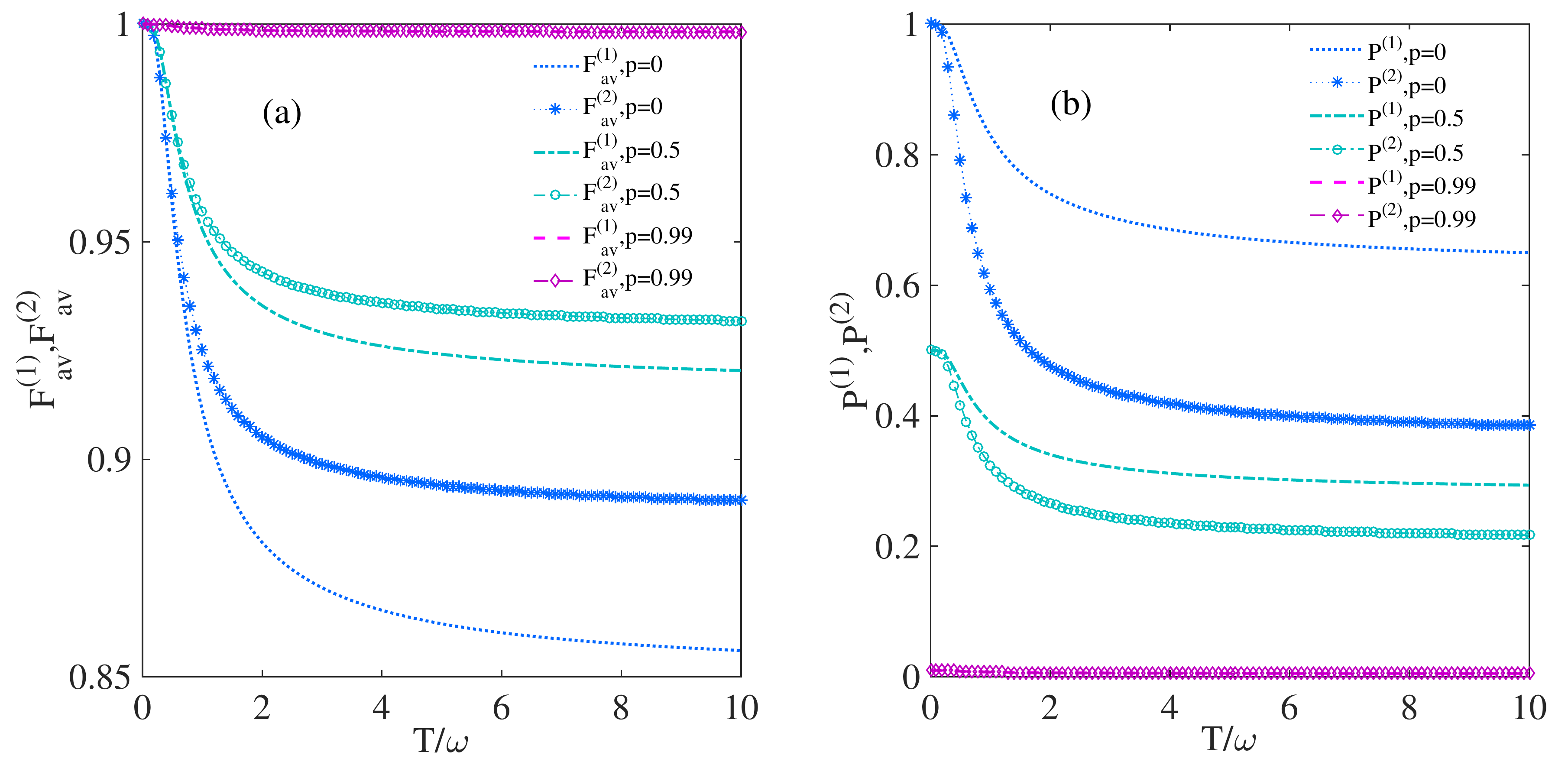}
\caption{(color online) The average fidelities $F_{\rm av}^{(1)}$, $F_{\rm av}^{(2)}$ and (b) the success probabilities $P^{(1)}$, $P^{(2)}$ as a function of scaled Hawking temperature for various values of $p$.}
\label{Fig3}       
\end{figure*}

The second method to find the proper post-weak measurement is calculating the following conditions:
\begin{equation}
\label{eq30}
\frac{\partial F_{\rm av}}{\partial\overline{q}}=0,\ \frac{\partial^2 F_{\rm av}}{\partial\overline{q}^2}<0.
\end{equation}
Then we obtain another optimal strength of the post-weak measurement
\begin{equation}
\label{eq31}
\overline{q}^{(2)}=\frac{\overline{p}\zeta^2\left(\sqrt{r^4+r^2+1}-r^2\right)^2}{\left(1+r^2\right)^2}
\end{equation}
with $r=\sqrt{\overline{p}}\eta$. With the above equation in hand, we can calculate the corresponding average fidelity $F_{\rm av}^{(2)}$. Here, we do not plan to present the analytic expression since it is too complicated. However, the Fig.~\ref{Fig3}a depicts the fidelities $F_{\rm av}^{(2)}$ (thick lines) as a function of Hawking temperature. For comparison, we also plot the fidelities $F_{\rm av}^{(1)}$ (thin lines) in the same figure. It is remarkable to notice that the later type of post-weak measurement always outperforms the former one given the same strength of pre-weak measurement. However, we should recognize that this superiority is based on the lower success probability, which is clearly shown in Fig.~\ref{Fig3}b.

\section{Discussions}
\label{sec4}
The underlying physical mechanism of the enhancement of fidelity needs to be clarified further. It is generally believed that the teleportation fidelity is an operational measure of quantum entanglement. Thus, it is expected that the more entanglement shared between Alice and Bob, the higher the fidelity will be in the standard teleportation protocol. This mechanism lies at the heart of the noisy quantum teleportation, and also in the relativistic quantum teleportation since the Unruh-Hawking radiation is a type of noise from the perspective of quantum information theory. In the absence of weak measurements, previous studies have unambiguously suggested that the Unruh-Hawking radiation will lead to the redistribution of entanglement between Alice, Bob (in region I) and anti-Bob (in region II) \cite{wang2010,emm2011}. Consequently, due to the monogamy of entanglement, this inevitably results in the entanglement degradation between the teleportation partners and thus a loss of the fidelity. 
Naturally, we wonder whether the enhancement of fidelity is still dependent on the increase of shared entanglement
when pre- and post-weak measurements are introduced.
 
 \begin{figure}[b]
 \includegraphics[width=0.45\textwidth]{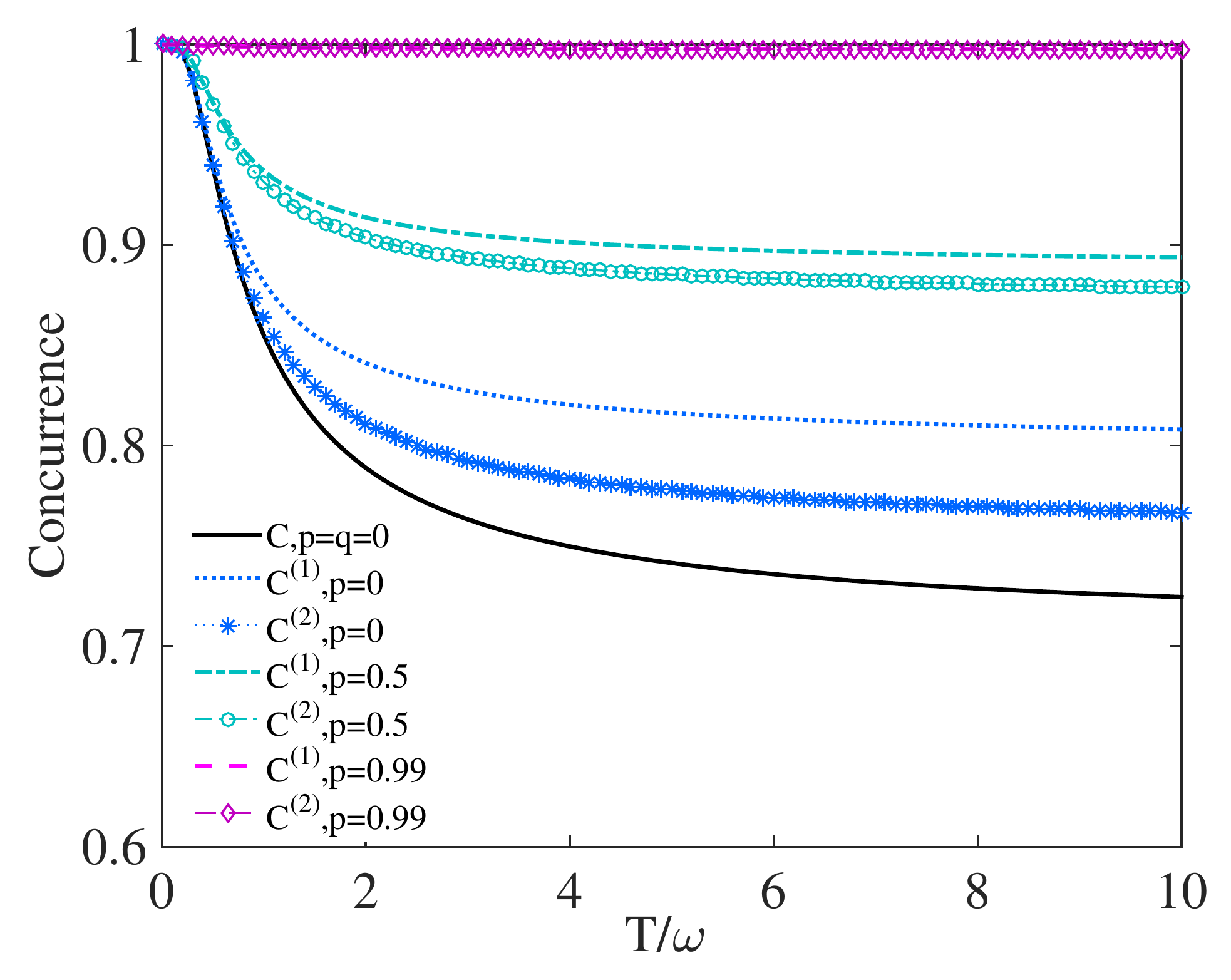}
\caption{(color online) The concurrences $C^{(1)}$ and $C^{(2)}$ as a function of scaled Hawking temperature for various values of $p$.}
\label{Fig4}       
\end{figure}
 
To answer this question, let us check the behaviours of entanglement in the presence of weak measurements.
By tracing over the other degrees of freedom, the reduced density matrix between Alice and Bob is given as
\begin{eqnarray}
\label{eq32}
\rho_{\rm A,I}=\frac{1}{N}\left(\begin{array}{cccc}\overline{p}\zeta^2 & 0 & 0 & \sqrt{\overline{pq}}\zeta \\0 & \overline{pq}\eta^2 & 0 & 0 \\0 & 0 & 0 & 0 \\\sqrt{\overline{pq}}\zeta & 0 & 0 & \overline{q}\end{array}\right).
\end{eqnarray}
The entanglement of $\rho_{\rm A,I}$, qualified by concurrence \cite{wootters1998}, is calculated as 
\begin{equation}
\label{eq33}
C=2\sqrt{\overline{pq}}\zeta/N.
\end{equation} 
Notice when $p=q=0$, then $N=2$ and the Eq. (\ref{eq33}) reduces to the concurrence between Alice and Bob in the background of Schwarzschild space-time without involving weak measurements.
On the basis of Eq. (\ref{eq27}), the optimal concurrence is given by
\begin{equation}
\label{eq34}
C^{(1)}=\frac{2}{2+\overline{p}\eta^2}.
\end{equation} 

\begin{figure*}
 \includegraphics[width=0.9\textwidth]{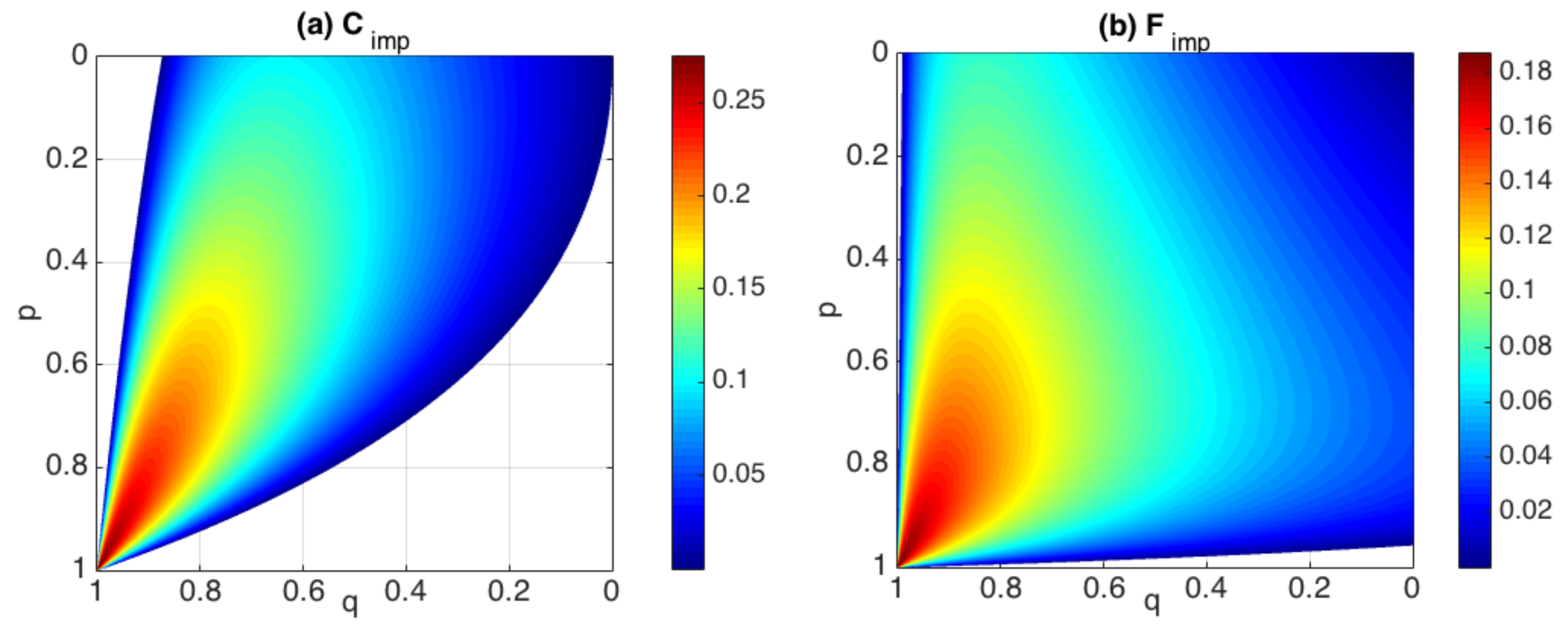}
\caption{(color online) The contour plots of (a) concurrence improvement $C_{\rm imp}$ and (b) fidelity improvement $F_{\rm imp}$ as a function of $p$ and $q$ with $T=10\omega$.}
\label{Fig5}       
\end{figure*}

Following the same procedure, we can also obtain the expression of $C^{(2)}$ by substituting Eq. (\ref{eq31}) into Eq. (\ref{eq33}). The behaivours of $C^{(1)}$ and $C^{(2)}$ as a function of Hawking temperature for various values of $p$ are presented in Fig.~\ref{Fig4}. Remarkably, Fig. \ref{Fig4} indicates that both $C^{(1)}$ and $C^{(2)}$ are larger than the concurrence in the absence of weak measurements (i.e., $p=q=0$). Thus, one might deduce that the enhancement of teleportation fidelity originates from the improvement of entanglement. However, a careful comparison between Figs.~\ref{Fig3}a and \ref{Fig4} reveals a significantly surprising phenomenon: Given the fixed value of $p$, even though the average fidelity $F_{\rm av}^{(2)}$ is greater than $F_{\rm av}^{(1)}$, the concurrence $C^{(2)}$ may be less than $C^{(1)}$. This result definitely implies that the the enhancement of teleportation fidelity cannot be straightforwardly attributed to the improvement of entanglement since the more entanglement does not mean higher fidelity.

In order to achieve a better understanding of this anomalous phenomenon, we introduce the definitions of concurrence improvement $C_{\rm imp}$ and fidelity improvement $F_{\rm imp}$,
\begin{eqnarray}
\label{eq35}
C_{\rm imp}&\equiv&C(p,q)-C_0,\\
\label{eq36}
F_{\rm imp}&\equiv&F(p,q)-F_0,
\end{eqnarray} 
where $C_0=\zeta$ and $F_0=(\zeta+1)^2/4$ are the concurrence and average fidelity without weak measurements. Then we can easily verify whether the fidelity enhancement is based on the entanglement improvement. The contour plots in Fig.~\ref{Fig5} show the valid regions of $p$ and $q$ for enhancing concurrence and fidelity. The blank areas indicate $C_{\rm imp}$ or $F_{\rm imp}$ less than 0. It is intriguing to notice that the valid regions in Fig.~\ref{Fig5}a and \ref{Fig5}b are inconsistent. This means that the fidelity enhancement could be realized even though the shared entanglement decreases. 
Therefore, we claim that the enhancement of teleportation fidelity could not be attributed to the improvement of entanglement, but rather to the probabilistic nature of weak measurements.

It seems that our scheme is the same as probabilistic teleportation \cite{li2000,pati2002}, whereas the physical mechanisms behind are completely different. The standard probabilistic teleportation is realized when the condition of ``entanglement matching'' is satisfied, while our method depends on the combination of pre- and post-weak measurements. Moreover, our scheme is still effective for the noisy teleportation even under severe decoherence.
There is also a clear similarity between the approach presented here and the procedure of Procrustean method of entanglement concentration which focus on a single copy of a partially entangled state \cite{bennett1996}. In both cases a preposed local operation and a post-local operation are performed successfully with a certain probability. However, as we discussed above, it is the fidelity not the entanglement that is concentrated by these local operations in this paper.

\section{Conclusions}

\label{sec5}

In summary, we propose the enhancement of teleportation fidelity in the background of Schwarzschild spacetime utilizing weak measurements. The crucial ingredient for the working of this proposal is an intentional introduction of a pre-weak measurement which partially collapses Bob's state towards the $|1\rangle$ state because of its laziness to Hawking radiation.
Due to the reversibility of weak measurements, the fidelity of teleportation could be greatly protected from the Hawking radiation with the assistance of post-weak measurement. In addition, we demonstrate that the enhancement of fidelity is not relative to the maintenance of entanglement, but rather to the probabilistic nature of weak measurements.
Our findings suggest that the technique of weak measurements is not only favorable to various quantum information processing tasks, but also could be powerful in relativistic quantum information, particularly, when the observers are subject to Unruh-Hawking radiation.

\begin{acknowledgements}
This work is supported by the Funds of the National Natural Science Foundation of China under Grant Nos. 11665004, 11605166 and 11365011, and supported by Scientific Research Foundation of Jiangxi Provincial Education Department under Grants Nos. GJJ150996 and GJJ150682. YL Li is supported by the Program of Qingjiang Excellent Young Talents, Jiangxi University of Science and Technology.
\end{acknowledgements}



\end{document}